\documentclass{PoS}

\title{Quark number fluctuations at high temperatures}

\ShortTitle{Quark number fluctuations at high temperatures}

\author{Peter Petreczky$^a$, Prasad Hegde$^b$ and
\speaker{Alexander Velytsky}$^{,a}$ (RBC-Bielefeld 
Collaboration)\\
$^a$Department of Physics, Brookhaven National Laboratories, Upton, NY11733, USA.\\
E-mail: \email{petreczk@quark.phy.bnl.gov} \\
E-mail: \email{vel@quark.phy.bnl.gov}\\
$^b$Dept. of Physics and Astronomy, SUNY Stony Brook, Stony Brook, NY11790, USA.\\
E-mail: \email{phegde@quark.phy.bnl.gov}
}

\abstract{
We calculate the second, fourth and sixth order quark number
fluctuations in the deconfined phase of 2+1 flavor QCD
using lattices with temporal extent $N_t=4, 6, 8$ and 12. We consider light, strange and
charm quarks. We use p4 action for valence quarks and gauge configurations
generated with p4 action with physical value of the strange
quark mass and light quark mass $m_q=0.1m_s$ generated by the RBC-Bielefeld
collaboration.
We observe that for all quark masses the quark number fluctuations rapidly
get close to the corresponding ideal gas limits. We compare our results
to predictions of a quasi-particle model and resummed high temperature perturbative
calculations. We also investigate correlations among different flavor
channels.
}

\FullConference{The XXVII International Symposium on Lattice Field Theory\\
		 July 26-31, 2009\\
		 Peking University, Beijing, China}

\begin{document}

\section{Introduction}
Quark number fluctuations are basic observables which can be obtained by taking 
derivatives of
the grand canonical potential with respect to quark chemical potentials $\mu_q$. 
Here we look directly at the related derivatives $c^q_i$, which 
enter into the Taylor expansion of the pressure
at finite chemical potentials ($\mu_q/T\lesssim 1$)
\begin{equation}
\frac{p}{T^{4}}=\frac1{VT^3}\ln Z(V,T,\mu_f)
=\left.\frac{p}{T^{4}}\right|_{\mu=0}+
\sum_{i,j,k,l}c^{u,d,s,c}_{i,j,k,l}(T)\left(\frac{\mu_{u}}{T}\right)^{i}
\left(\frac{\mu_{d}}{T}\right)^{j}\left(\frac{\mu_{s}}{T}\right)^{k}
\left(\frac{\mu_{c}}{T}\right)^{l}.
\label{eq:PTaylor}
\end{equation}

Their combinations form fluctuations of 
conserved charges (see ref. \cite{Cheng:2008zh} for a related lattice study).
Thus they can be used to study the properties of a thermal medium produced in
heavy ion collisions \cite{Koch:2008ia}. From Eq. (\ref{eq:PTaylor}) 
it is clear that
the expansion coefficients are related to the quark number densities $n_f$, 
quark number susceptibilities $\chi^f_2\equiv\langle n^2_f\rangle=2c^f_2$ 
and other moments: 
$\chi^f_4\equiv\langle n^4_f\rangle -3\langle n^2_f\rangle^2=24c^f_4$, $\ldots$

In this work we are interested in the behavior of the quark number fluctuations at 
high temperatures $240{\rm MeV}<T<900{\rm MeV}$ and zero chemical potential. 
We would like to study the approach of relevant observables to the ideal gas
(Stefan-Boltzmann (SB) limit)
\cite{Allton:2005gk}
\begin{equation}
\frac{p_{SB}}{T^4} = \Omega^{(0)} (T,\mu) =  \frac{8 \pi^2}{45} + 
 \sum_{f=u,d,..} \left[\frac{7 \pi^2}{60} +
\frac{1}{2}  \left(\frac{\mu_f}{T}\right)^2 
+ \frac{1}{4 \pi^2} \left(\frac{\mu_f}{T}\right)^4 
\right].
\label{eq:sb}
\end{equation}

We use the tree level Symanzik improved gauge action 
and improved staggered fermion p4-action with 3-link smearing (p4fat3). The
gauge configurations were generated for calculation of the equation of state 
with $(2+1)$ flavors using almost physical light quark masses resulting in 
$m_\pi\approx 220MeV$ and physical $s$ quark mass \cite{Cheng:2007jq}.
In our measurements we add non-dynamical $c$ quark. In table
\ref{tab:conf} we list the gauge configurations used in the study.

\begin{table}[ht]
\small
\begin{tabular}{c|c|c|c|c||c|c|c|c|c}
lattice&$\beta$& $T$ & configurations&Rnd&
lattice&$\beta$& $T$ & configurations&Rnd\\\hline
$32^3\times 8$&4.08&539.19&24100$^*$&96 &$32^3\times 4$&3.92&861& 6500&96\\
$32^3\times 8$&4.00&474.84&23200$^*$&96 &$16^3\times 4$&3.82&706& 10000&96\\
$32^3\times 8$&3.92&416.42&27100$^*$&96 &$16^3\times 4$&3.76&622& 11650&96\\
$32^3\times 8$&3.82&350.66&15100$^*$&96 &$16^3\times 4$&3.69&532& 9450&96\\
$32^3\times 8$&3.76&314.53&6050&96 &$16^3\times 4$&3.63&465& 10000&96\\
$32^3\times 8$&3.63&242.78&6000&96 &$16^3\times 4$&3.57&404  & 21150&96\\
&&&&&$16^3\times 4$&3.54&373  & 6250&96\\
&&&&&$16^3\times 4$&3.51&347  & 10000&96\\
&&&&&$16^3\times 4$&3.49&330  & 9400&96\\
\hline
$32^3\times 6$&4.08&745&6800&96& $32^3\times12$&3.76&212&3150&100\\
$32^3\times 6$&4.00&633&10000$^*$&96& $32^3\times 12$&3.82&240&3750&100\\
$32^3\times 6$&3.92&553&9300&96& $32^3\times 12$&3.92&273&4000&100\\
$32^3\times 6$&3.82&451&8400&96\\
\hline
\end{tabular}
\caption{\label{tab:conf}
Lattice parameters and the number of trajectories for each parameter set. The step size between measurements 
is 100 trajectories
for stared entries ($^*$) and 50 for the rest. 
Typically we start measurements after 1000 trajectories. 
Rnd is the number of random vectors used in the stochastic estimators.
}
\end{table}

\section{Numerical results}
\begin{figure}[ht]
\begin{center}
\includegraphics[width=0.49\textwidth]{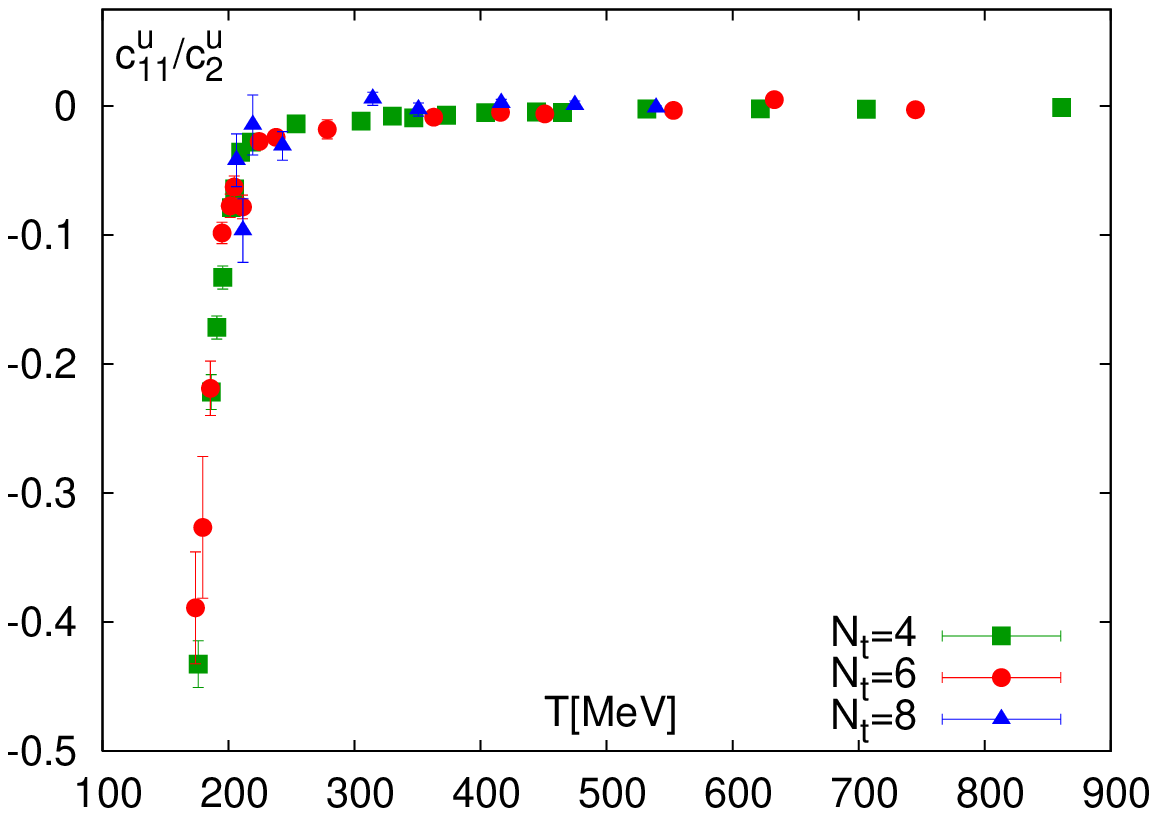}
\includegraphics[width=0.49\textwidth]{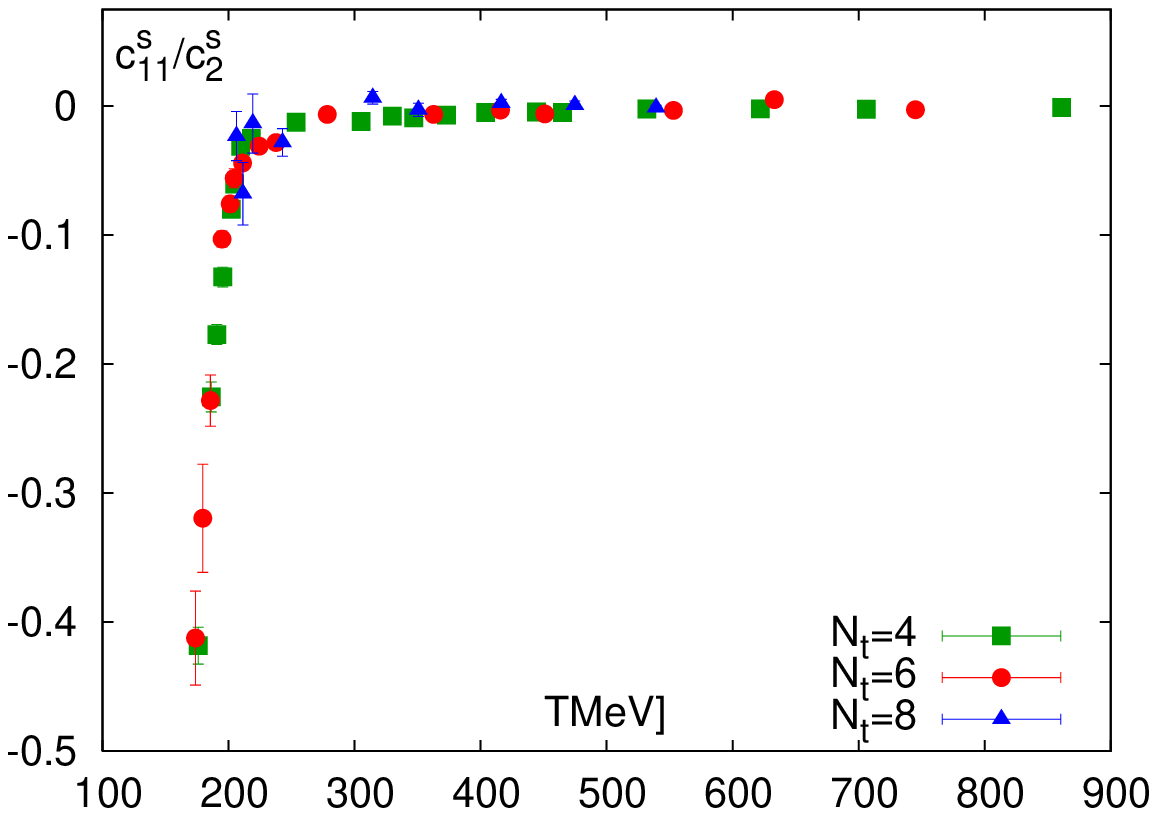}
\end{center}
\caption{\label{fig:11}The off-diagonal coefficient $c^{ud}_{11}$ (left) and 
$c^{us}_{11}$ (right) normalized by $c^u_2$ and $c^s_2$ respectively.
}
\end{figure}

Here we present our numerical results for quark number fluctuations up to the 6th order,
including second order flavor off-diagonal fluctuations.
In Fig. \ref{fig:11} we plot the off-diagonal coefficients 
$c^{ud}_{11}$ and $c^{us}_{11}$ 
normalized by the related diagonal quadratic terms $c^u_2$ and $c^s_2$. The plot indicates that 
the correlations between $u$ and $d$, and $u$ and $s$ flavors vanish rapidly after the deconfinement.
This is suggestive of the fact that there are no bound states in the deconfinement phase
\cite{Koch:2005vg}.
Similar results were obtained for $uc$ correlations. 
Thus we may expect that quark gas description should be reasonable for quark number
fluctuations.

\begin{figure}[ht]
\includegraphics[width=0.49\textwidth]{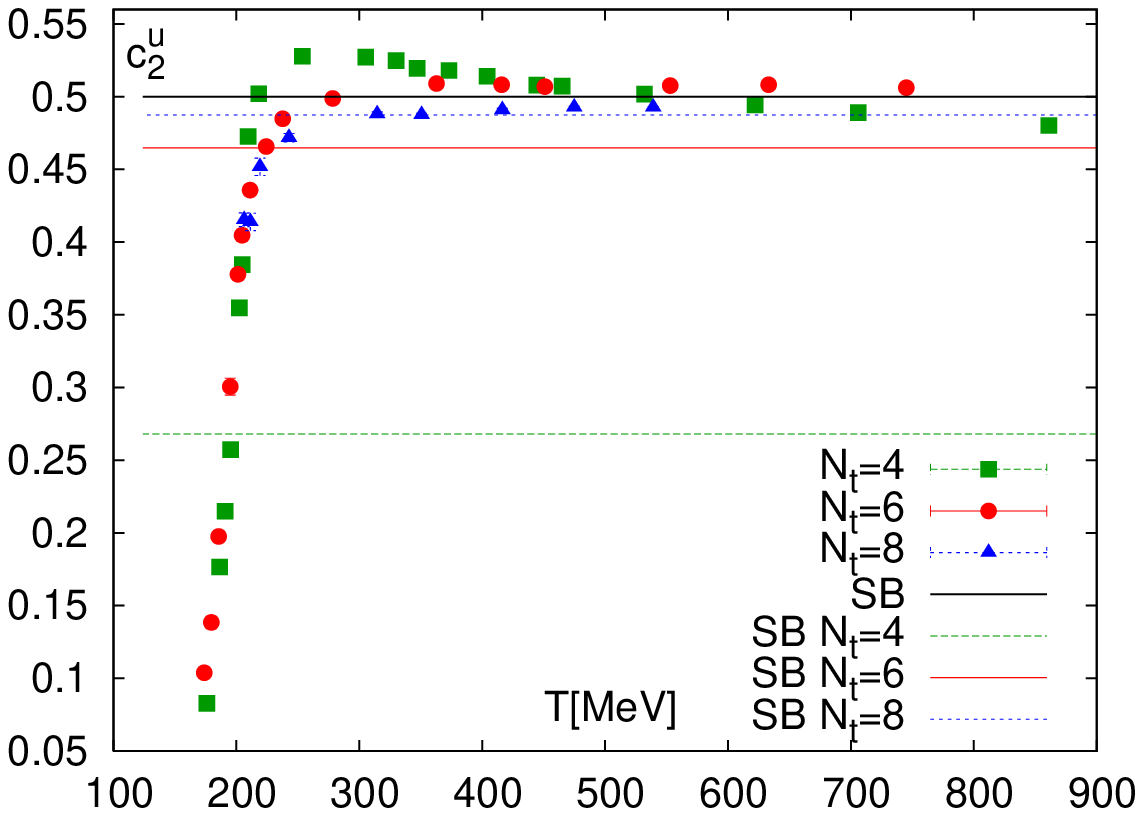}
\includegraphics[width=0.49\textwidth]{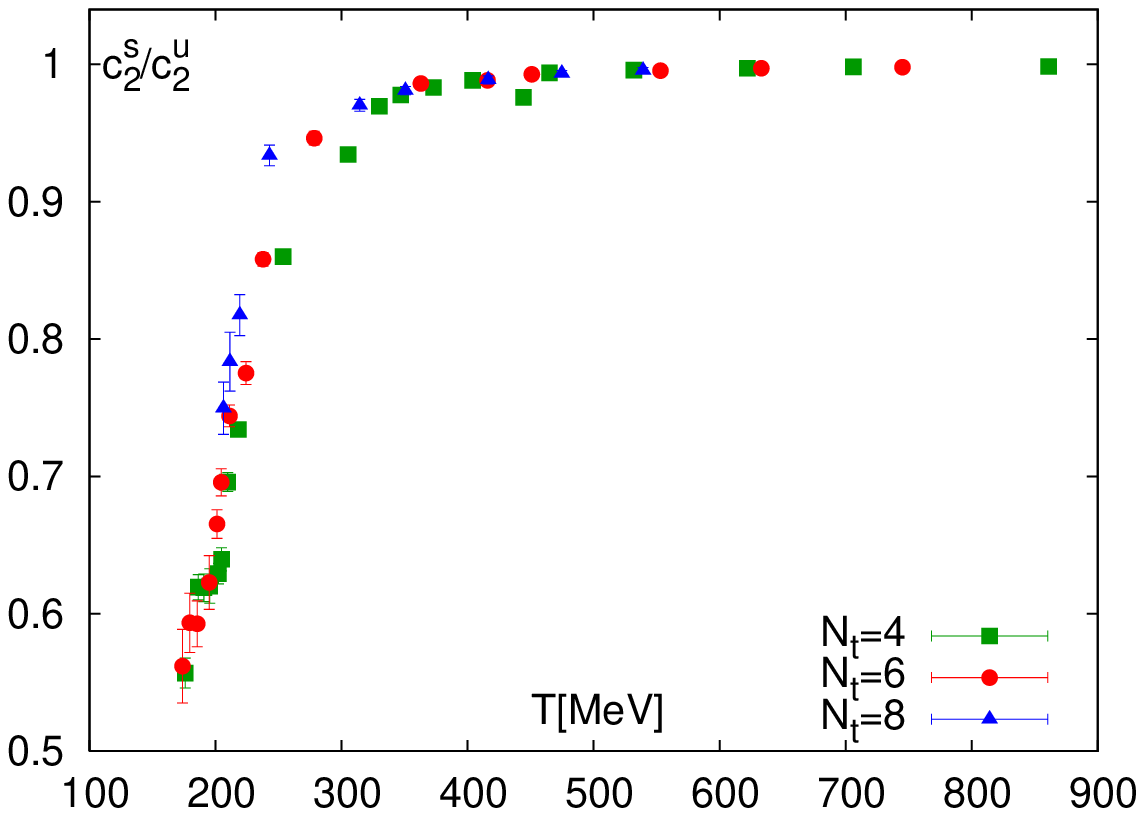}
\caption{\label{fig:c2u} Numerical results for $c_2^u$ calculated for different $N_t$ 
and the corresponding SB values (left). The ratio 
$c_2^s/c_2^u$ as function of the temperature (right).}
\end{figure}
In the left panel of Fig. \ref{fig:c2u} we present our lattice 
results for $c_2^u$ coefficients as well as the corresponding
continuum SB limit.
In addition we plot the SB value corrected for a finite lattice cutoff.
In Fig. \ref{fig:c2u} we also show our previous (low temperature) results from Ref. \cite{Cheng:2008zh}.
We see that the observed cutoff effects are different and smaller than the
cutoff effects in the free theory. The difference between the $N_t=6$ and $N_t=8$
results is about $10\%$. Therefore to establish the continuum limit calculations 
on $N_t=12$ lattices are needed.
Similar results were obtained for $c_2^s$. The right panel of Fig. \ref{fig:c2u}
shows the ratio $c_2^s/c_2^u$. As this has been discussed in \cite{Cheng:2008zh,
Bazavov:2009zn} this ratio approaches unity at $T\geq300$MeV. This in fact
is expected in the quark gas picture.

\begin{figure}[ht]
\includegraphics[width=0.49\textwidth]{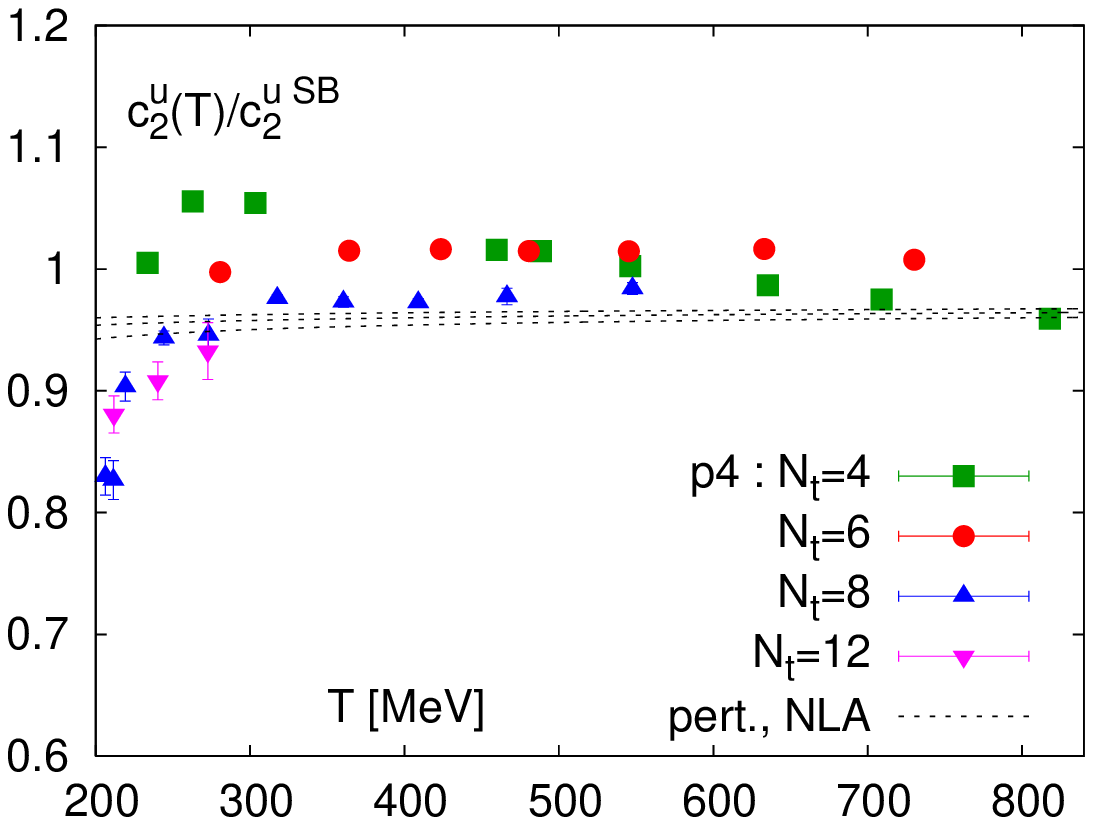}
\includegraphics[width=0.49\textwidth]{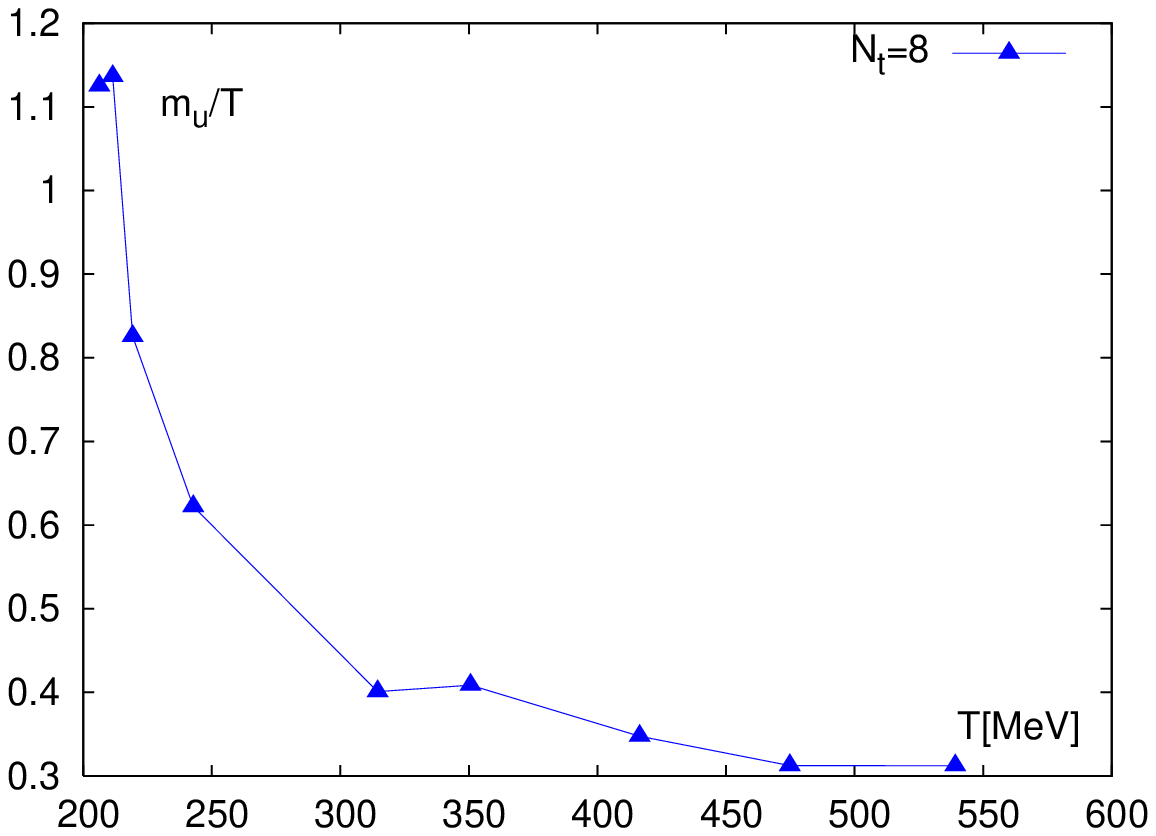}
\caption{\label{fig:cu} Left panel: The ratio of $c_2^u$ to the corresponding SB 
value compared to the 
high temperature perturbative results shown as dashed lines. The three lines
correspond to the renormalization scale $\mu=\pi T, 2\pi T$ and $4\pi T$ (from
bottom to top).
Right panel: The effective thermal quark mass extracted from $c_2^u$. 
}
\end{figure}

In the left panel of Fig. \ref{fig:cu} we compare our $c_2^u$ results for $T\geq200$MeV with 
the resummed 
high temperature perturbative calculations
\cite{Rebhan:2003fj}. 
This figure clearly shows the $10\%$ difference between $N_t=6$ and $N_t=8$ results.
Therefore we also did calculations at $N_t=12$ lattices at 3 temperatures.
The corresponding results are also shown in the figure. At the highest temperature,
where $N_t=12$ data are available,  we see a reasonably good agreement with $N_t=8$
data. This suggests that $N_t=8$ calculations are sufficiently close to the 
continuum results. Therefore the comparison of $N_t=8$ data with the resummed
perturbative calculations performed in the continuum limit is meaningful and shows
that quark number fluctuations are well described by the resummed perturbative calculations.

\begin{figure}[ht]
\begin{center}
\includegraphics[width=0.75\textwidth]{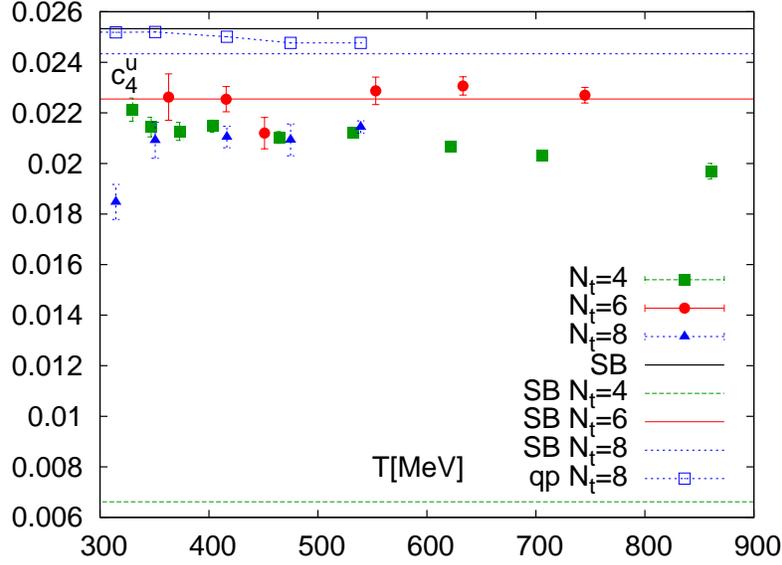}
\caption{\label{fig:c4u} The high temperature behavior of $c_4^u$. }
\end{center}
\end{figure}

We also calculated the 4th order expansion coefficients $c_4^u$ and $c_4^s$ in
the high temperature regime. The numerical results for $c_4^u$ are shown in 
Fig. \ref{fig:c4u} (the results for $c_4^s$ are similar at high
temperature).
In the figure we show the corresponding SB value in the continuum 
as well as for finite $N_t$. The cutoff dependence of $c_4$ is similar to that 
of $c_2$. Namely, the observed cutoff effects are different from the free
theory. Also the ordering of $N_t=6$ and $N_t=8$ results is reversed. The difference
between them is less than $10\%$. Assuming that $N_t=8$ results are close to the 
continuum limit, as the $c_2$ calculations suggest, we see deviations from
the ideal gas limit around $20\%$, i.e. two times more than for $c_2$.

\begin{figure}[ht]
\includegraphics[width=0.49\textwidth]{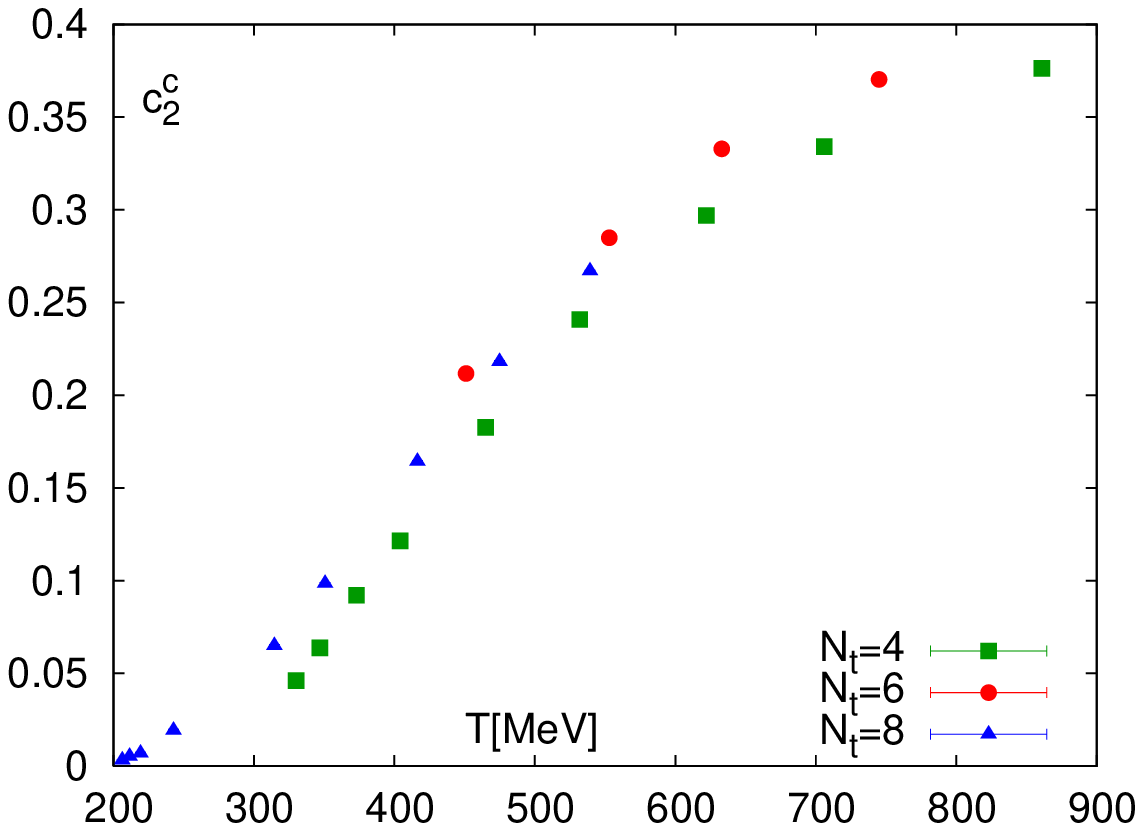}
\includegraphics[width=0.49\textwidth]{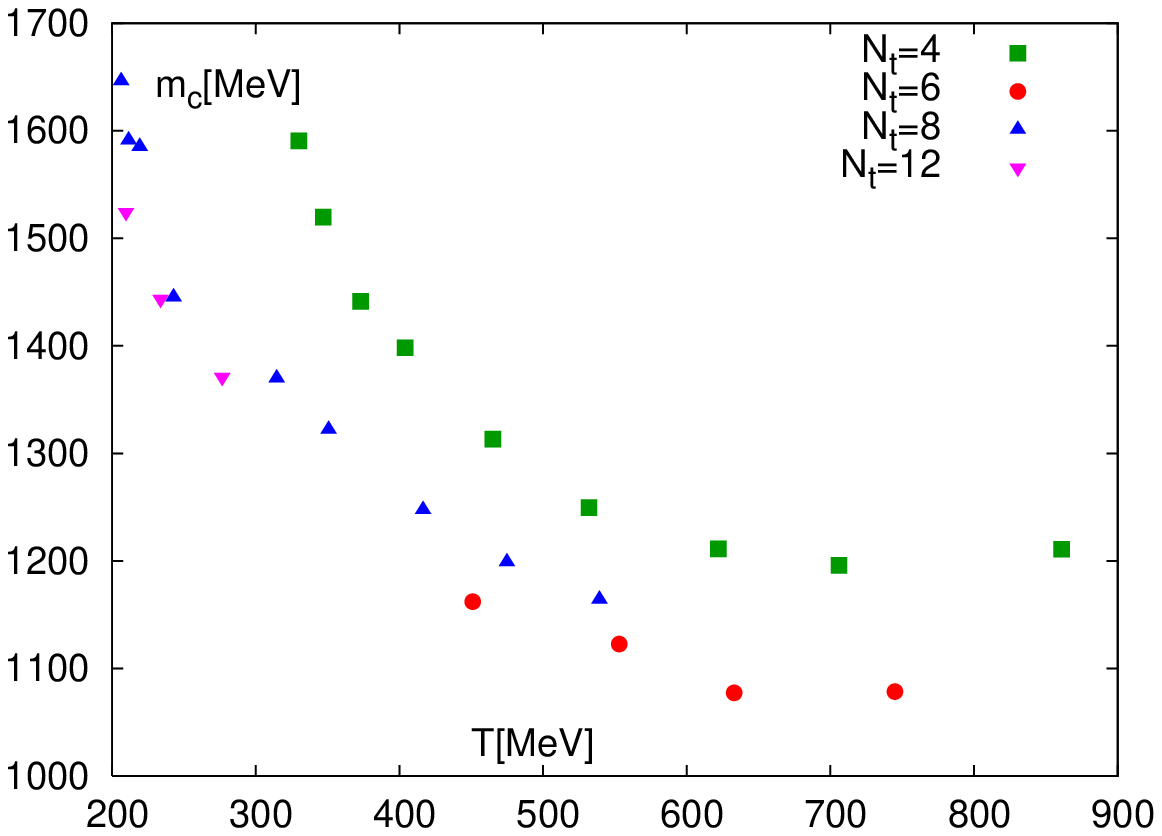}
\caption{\label{fig:c2_4c} Numerical results for $c_2^c$ (left)
 and the effective mass extracted from it (right).}
\end{figure}
In the resummation scheme used in \cite{Rebhan:2003fj} the dominant contribution 
to the quark number fluctuations is given by quasi-particles with effective quark mass
equal to the hard thermal loop mass. Therefore it makes sense to compare the lattice
results to a simple massive quasi-particle model.
The pressure for massive quark gas is given by equation (\ref{eq:pm}) 
\cite{Karsch:2003zq}
\begin{eqnarray}
\label{eq:pm}
\frac{p_m}{T^4}&=&\frac{6}{\pi^2}\left(\frac m T\right)^2\sum_{l=1}^{\infty}(-1)^{l+1}l^{-2}K_2(lm/T)\cosh
(l\mu_q/T),\\
\label{eq:c2qp}
c_2        &=& \frac12\frac{6}{\pi^2}\left(\frac m T\right)^2\sum_{l=1}^\infty (-1)^{l+1} 
                 K_2(lm/T).
\end{eqnarray}
Taking successive derivatives of the rhs with respect to $\mu/T$ we obtain
expressions for $c_2$ (\ref{eq:c2qp}), $c_4$, etc. The expression for $c_2$ can be solved for $m/T$
using $c_2$ values obtained numerically. 
The corresponding quasi-particle masses are shown in Fig. \ref{fig:cu}
(right) for $N_t=8$. We can use these effective masses to make predictions for 
$c_4$ and $c_6$. The quasi-particle results for $c_4^u$ is shown in Fig. \ref{fig:c4u}.
As we can see they are close to the massless ideal gas limit and almost $20\%$
off from the numerical data for $N_t=8$.
This indicates that the simplest quasi-particle description
is not adequate for light quarks. Unfortunately the are no perturbative calculations for $c^u_4$.
It would be interesting to see whether the resummed perturbative calculations similar to those
performed for $c^u_2$ in \cite{Rebhan:2003fj} could explain the observed deviation from the ideal
gas limit for $c^u_4$.

Now we consider the charm quark case.
In Fig. \ref{fig:c2_4c} on the left panel we plot the results for $c_2^c$. Here one can notice that $N_t=4$ data shows slightly different
scaling due to cutoff effects. On the right panel we present the effective 
mass extracted from the $c_2^c$ values at each temperature point. 
\begin{figure}[ht]
\begin{center}
\includegraphics[width=0.75\textwidth]{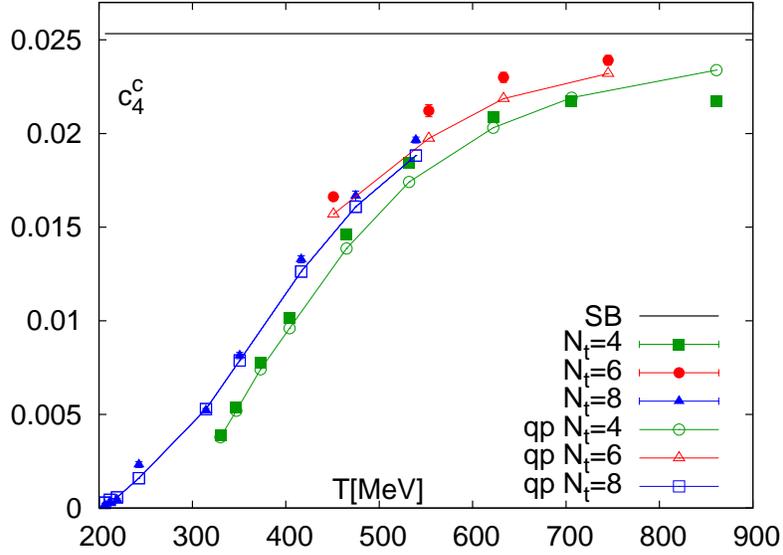}
\caption{\label{fig:c4c} Numerical results for $c_4^c$ and the quasi-particle results.}
\end{center}
\end{figure}
In Fig. \ref{fig:c4c} we compare $c_4^c$ data with the quasi-particle model results, where the effective mass was computed from
$c_2^c$ data. We see a very good agreement between the quasi-particle model results and
the numerical data. This indicates that for heavy charm quark the quasi-particle
description is adequate for modeling of $c_4^c$ at high temperature.

\begin{figure}[ht]
\begin{center}
\includegraphics[width=0.75\textwidth]{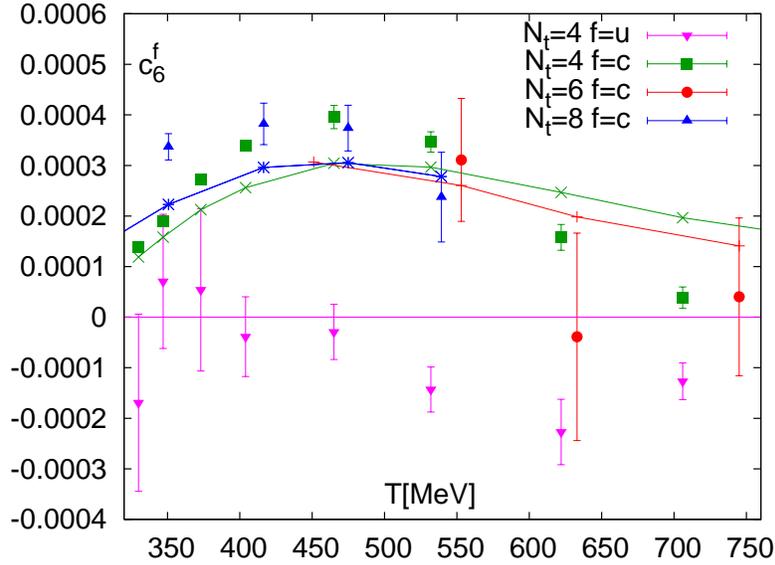}
\caption{\label{fig:c4s} The sixth order $c_6$ results for $c$ and $u$ flavors.}
\end{center}
\end{figure}

As we see from Figs. \ref{fig:c2_4c} and \ref{fig:c4c} the fluctuations of the charm quark
number is comparable to those of the light quark sector at the highest temperatures. This
means that at temperatures $T\sim 800$MeV the contribution of charm quark to QCD thermodynamics
is significant as this has been suggested in Ref. \cite{Laine:2006cp} using weak coupling
analysis.

We also considered the 6th order coefficient.
In Fig. \ref{fig:c4s} we present the results for $c_6^u$ and $c_6^c$.
The light quark result according to Eq. (\ref{eq:sb}) should be zero for a massless
ideal gas. The numerical result agrees reasonably with this prediction. For the 
charm quark number fluctuation
$c_6^c$ we took the effective mass computed from $c_2^c$ and used it to compute 
the massive ideal gas estimate, which we plot in the figure with lines for various
$N_t$. It is interesting that the numerical results for all $N_t$ 
are in reasonable agreement with the quasi-particle model results in the 
deconfined regime.

\section{Conclusions}
In this contribution we discussed quark number fluctuations $c_2, c_4$ and $c_6$ 
calculated with p4 action on $N_t=4,6,8$ and 12 lattices. We see that the cutoff 
dependence of quark number fluctuations is different from the expectation
based on the free theory. Off-diagonal fluctuations vanish soon after the deconfinement
transition. This suggests that fluctuations can be understood in terms of the quark
gas. In the light quark sector the quadratic fluctuations are quite close to 
the massless ideal gas limit and agree with the resummed perturbative calculations
and expectations of the quasi-particle model. The quartic fluctuations
on the other hand show larger (about $20\%$) deviation from the massless ideal gas 
limit and do not agree with the predictions of the quasi-particle model. 
In the charm quark sector we find that all fluctuations up to 6th order can be 
reasonably well described by the quasi-particle model. Current results suggest
that continuum limit may be reached for $N_t=8$ but calculations on $N_t=12$
lattices at higher temperature are needed to verify this statement.

\section*{Acknowledgments}
This work was supported by U.S. Department of Energy under Contract No. DE-AC02- 
98CH10886.

\end{document}